\documentclass[11pt,twocolumn]{article}

\newfont{\mycrnotice}{ptmr8t at 7pt}
\newfont{\myconfname}{ptmri8t at 7pt}
\usepackage[a4paper,bindingoffset=0.2in,%
            left=0.5in,right=0.5in,top=0.8in,bottom=0.8in,%
            footskip=.25in]{geometry}

\usepackage[caption=false]{subfig}
\usepackage{bbm}
\usepackage{multirow}
\usepackage{xcolor}
\usepackage[hyphens]{url}
\usepackage{lipsum}
\usepackage{amsfonts,amsmath,tikz,hyperref,textcomp}
\usepackage{etoolbox}
\usepackage{multicol}
\usepackage{adjustbox}
\graphicspath{{}{figures/}}

\begin{document}

\title{Probabilistic Modelling of the Impact on Bus Punctuality of a Speed Limit Proposal in Edinburgh (Extended Version)}

\author{
  Dani\"el Reijsbergen \;\;\;\;\;\;\;\;\;\;
  Rajeev Ratan
}

\twocolumn[
  \begin{@twocolumnfalse}
    \maketitle
  \end{@twocolumnfalse}
  ]

		\begin{abstract}
		We propose a data-driven methodology for evaluating the impact of the introduction of a speed limit on the punctuality of bus services. In particular, we use high-frequency Automatic Vehicle Location data to parameterise a model that represents the movement of a bus along predefined patches of the route. We fit the probability distributions of the time spent in each patch to two classes of probability distributions: hyper-Erlang distributions, for which we use the tool HyperStar, and a variation of the three-parameter gamma distributions recommended by the Traffic Engineering Handbook. In both cases we obtain models that can be expressed using the framework of Probabilistic Timed Automata, allowing us to evaluate bus punctuality using the model checking tool UPPAAL. We conduct a case study involving a proposed speed limit in Edinburgh. This is an extended version of a paper presented at ValueTools 2015 \cite{reijsbergen2015probabilistic}.
		\end{abstract}

%
%


\section{Introduction}

\begin{figure}[t]
	\framebox{\includegraphics[width=0.46\textwidth]{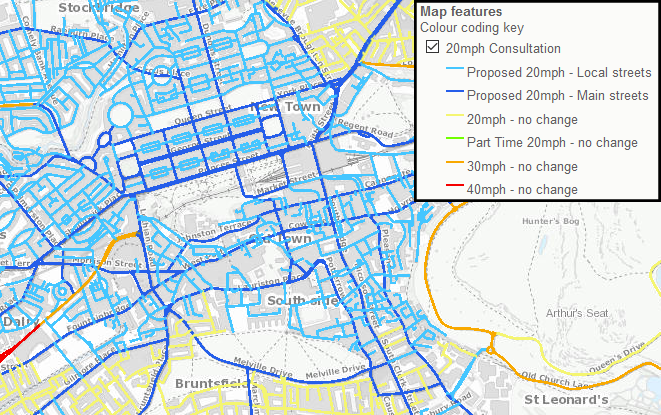}}
    \caption{Map of proposed 20mph streets, zoomed in slightly to focus on Edinburgh's city centre (blue and dark-blue roads will be changed into having 20mph speed limits). 
		An interactive version is available at \url{http://edinburghcouncilmaps.info/LocalViewExt/Sites/20mphConsultation/}
		}
    \label{fig: speed limit map}
\end{figure}

Anticipating the possible impact of changes in the operating environment on mass transit timetables has been a long-standing challenge for planners and operators. A present manifestation of this situation is seen in the City of Edinburgh, where the city council has agreed to implement a new speed limit along the majority of the city's roads after years of campaigning by road safety groups. The concerned roads are heavily utilised by Lothian Buses, the main urban bus company that operates in the City of Edinburgh, who pride themselves on maintaining a high level of service quality and reputation. Consequently, the delays emanating from the introduction of these speed limits have the possibility to adversely affect timetable adherence, resulting in missed targets set by regulating bodies, and an inconvenienced travelling public. The proposed speed limits (see Figure \ref{fig: speed limit map}) affect the majority of their routes currently operated. As such, quantifying the possible impacts of these speed limits is of major concern. 

In simulating the effects of changes to real world systems, there are two basic approaches. The first involves direct experimentation on the existing system under the new conditions (or a close approximation thereof). This, if done correctly, produces a realistic measure of the effect. However, many real world systems do not lend themselves to such direct experimentation as it is either costly or logistically infeasible. As such, the second approach seeks to create a model to mimic the relevant behaviour of the system. This model can then be used to simulate a variety of conditions or constraints. However, this poses a challenge as real world bus systems are often subject to stochastic behaviour, thus necessitating a suitable model to reflect this.

In this paper we perform a case study on two bus routes in Edinburgh and provide a quantitative answer to the question of how the proposed speed limit impacts bus punctuality. Our methodology for developing bus models involves an expanded and modified implementation of the work done in \cite{reijsbergen2015patch} and \cite{vissat2015finding}. Using this approach, we build a model that represents the continuous-time movement of buses as they move along predefined segments of a route. The time needed to move from one of the segments (referred to as `patches' throughout this paper) to the next is modelled as a random variable whose distribution is chosen to best fit high-frequency (approximately every 5 seconds) GPS data provided to us by Lothian Buses. In this paper, we compare two classes of distributions used to model the time in each of the patches. The first of these is the class of hyper-Erlang distributions, a modelling class with good performance in a broad range of settings and which is easily parameterised using the user-friendly tool HyperStar \cite{reinecke2012hyperstar}. The second is a variation of the three-parameter gamma distribution, which is an industry standard recommended by the Traffic Engineering Handbook \cite{pline1992traffic}; its parameters are determined using a purpose-built optimisation tool written in the language of the statistical package R \cite{team2005r}. The resulting distributions are then used to build probabilistic models in the tool UPPAAL \cite{behrmann2004tutorial}, which facilitates the modelling formalism of Probabilistic Timed Automata and the simulation-based verification framework of Statistical Model Checking (SMC) \cite{david2011time}.

We apply this modelling approach to a case study involving two bus services. The first is the Airlink, a popular bus route that connects Edinburgh's city centre to the airport. Big delays may cause disaffection among passengers travelling to the airport. However, we find that the proposed speed limit has a small impact, namely between 7.61 and 24.3 seconds depending on the time and the direction. The second is the N11 night service, which is particularly affected by the new speed limit seeing as 50\% of its route is affected by the speed limit, and because buses are particularly able to speed up to over 20mph at night. In this case, average delays are still only about 32 seconds. Finally, we fully exploit the stochastic model and the SMC engine of UPPAAL to compute `on-time' probabilities for the final stops of the Airlink, namely the probability of the bus arriving at the stop between at most one minute early and five minutes late. 

The structure of this paper is as follows. In Section~\ref{sec: preliminaries} we discuss the context of our work, describe the dataset, and provide an overview of related research. In Section~\ref{sec: modelling} we give a formal description of the modelling and parameter fitting techniques used in this paper, and discuss the use of UPPAAL. In Section~\ref{sec: results}, we use the presented methodology to quantify the impact of the speed limit plan on the Airlink service. Section~\ref{sec: conclusions} concludes the paper.




\section{Model \& Preliminaries}
\label{sec: preliminaries}

\subsection{Motivation \& Background} \label{sec: background}

The speed limit plan discussed in this paper was approved by Edinburgh's City Council on 13 January 2015, a decision which followed three years of research and three months of public consultation on the part of the council \cite{edinburghcouncil2015about}. The speed limit will be introduced gradually starting in April 2016 \cite{bbc2015edinburgh}. The move has been hailed by campaign groups such as Living Streets Edinburgh \cite{lse2014response}, 20's Plenty For Us and Friends of the Earth Scotland. The arguments in favour of the speed limit include public support for the speed limit witnessed across several opinion polls, improved road safety, cycling and walking being encouraged leading to lower carbon emissions and health benefits, and a nicer and more attractive city overall. 

The move has not been entirely uncontroversial \cite{edinburghnews2015huge}, with a Facebook campaign called `Say no to 20mph' being started and a related petition drawing 2700 signatories \cite{stv2015why}. Opponents of the speed limit plan are generally skeptical of the suggested benefits and argue that the speed limit will encourage antisocial behaviour (such as aggressive tailgating), that it will be ineffective if it is not properly enforced, and that journey times will be adversely affected. Longer journey times also have an impact on bus timetable adherence, which is of specific interest to Lothian Buses due to its effect on passenger satisfaction and the fact that they have been fined $\pounds$10,500 in the past for non-compliance with punctuality targets \cite{scotsman2010capital}. The cause of the fine was that the impact of road works was overestimated, leading to buses running early with high probability. Accordingly, there is a demand for predictions of the impact of the speed limit to be as accurate as possible. 

As we demonstrate in this paper, the accuracy of these predictions --- particularly for timetable adherence probabilities --- can be affected by the modelling choice. In previous work \cite{reijsbergen2015patch}, the hyper-Erlang distribution was chosen to model the time spent in patches for several reasons, including its general applicability and the fact that the resulting models are Continuous-Time Markov Chains, for which many efficient analysis techniques exist. However, as we discuss later on, an important alternative, namely the probability distribution that is recommended for traffic engineers, may not yield a Markov chain, but the resulting model can still be expressed using the framework of Probabilistic Timed Automata, allowing us to use the tool UPPAAL \cite{behrmann2004tutorial} and its powerful stochastic simulation engine. We compare the two in this paper, not just with the intent to make the predictions of the speed limit impact as accurate as possible, but also to inform the construction of more general urban traffic models as part of the EU project QUANTICOL.

\subsection{Patch Structure}

Although the Automatic Vehicle Location (AVL) measurements supplied to us by Lothian Buses are roughly done in continuous space, we discretise our model by dividing the full route into ordered patches and consider the travel time through each patch. Since our case study does not require the same road-level detail as in \cite{reijsbergen2015patch}, the patches were chosen to be relatively large, namely between 500m and 2.5km.

\begin{figure*}[!htbp]
	\centerline{\framebox{\includegraphics[width=0.95\textwidth]{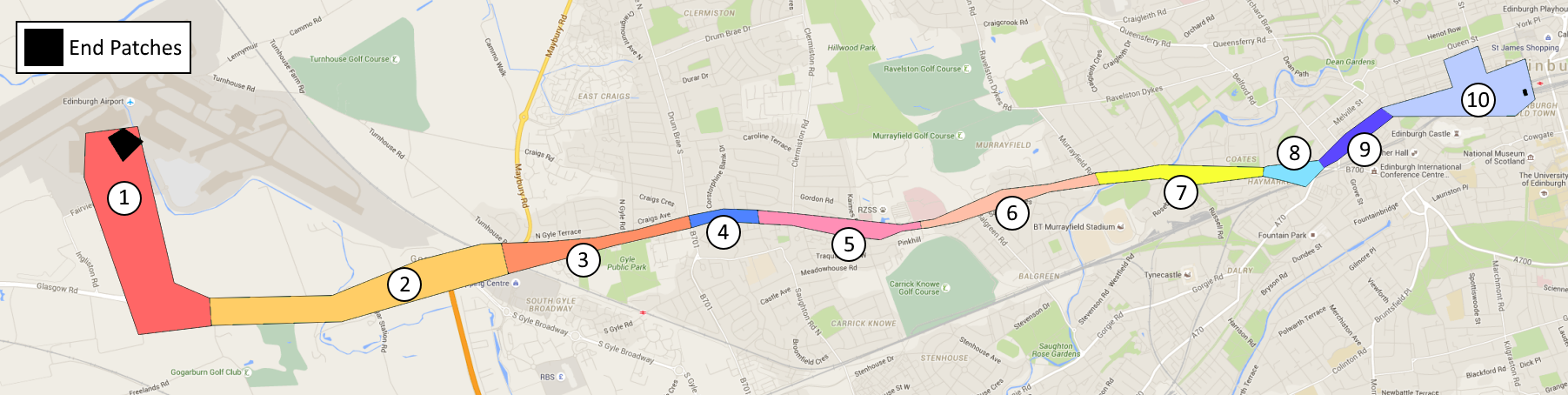}}}
    \caption{Illustration of the patches used to partition the Airlink route. Patches 4, 8, 9, and 10 will be affected by the new speed limit.}
    \label{fig: patches airlink}
\end{figure*}

Our patches were designed to encapsulate areas between major bus stops as shown in Figure \ref{fig: patches airlink}. Another important criterion was to ensure that the road segments covered by a patch had to either remain unaffected by the speed limit or be changed to 20mph, but not both. This was done so that we could implement the speed limit effects at a per-patch level. 
Finally, specific patches were constructed for the end points of the route, because buses tend to spend a large amount of time there; these are the black patches in Figure~\ref{fig: patches airlink}. We did not include the time spent in these patches in the total journey times.

\subsection{Description of Data}

The dataset used in this paper was gathered using the WiFi capabilities of some of the vehicles used by Lothian Buses. These WiFi services are provided via Long Term Evolution (LTE) 4G cellular modems which  allow for reliable high-frequency --- once every 5 seconds --- AVL measurements. It should be noted that non-WiFi equipped buses also relay AVL data as part of the MyBusTracker service that provides information for arrival time prediction at stops and web applications.\footnote{An API for the low-frequency data is publicly available via \url{http://www.mybustracker.co.uk/}.} However this is at a much lower frequency, namely between 30 to 40 seconds. For analysis of bus speeds, speed averages of over 40 seconds are typically too imprecise for accurate model parameterisation. 

The data set used for the case study contains information for all buses operating the Airlink route from Monday 20 July 2015 to Wednesday 22 July 2015. We also have access to data involving Route 31 and N11 night route vehicles from the same period, although we will only briefly consider the latter in Section~\ref{sec: results}. The data was provided in the form of files (in CSV format) containing all AVL measurements logged per bus per day, starting at 3AM in the morning and ending at 3AM the next day. Speed measurements were included in the dataset, although these could otherwise be calculated from the distance travelled.

Since traffic conditions vary throughout the day, we focus on two non-rush hour periods, namely the period between 9AM and 3PM and the period between midnight and 5AM. The reasons are twofold: even between individual hours in the rush hour period there are still considerable differences, and the impact of the speed limit is expected to be the largest outside the rush hours. We also categorise measurements based on the direction that is being travelled, as this may have an impact; in fact, the distance crossed in Patch~10 of Figure~\ref{fig: patches airlink} is not the same in both directions.

Routes are typically operated by several individual buses per day. In order to comply with the timetable, buses first leave the depot and then head towards the starting points of the route. Buses then leave these starting points in staggered intervals as per their timetable. This allows buses to service stops along the route in frequencies ranging from once per 5 to 15 minutes (60 minutes for the night bus). Within a single day (or night), buses may change route. This can be problematic, as patch crossing time measurements are inaccurate if only portions of a patch were completed. 
For the Airlink service this is not the case, because the vehicles of this service are painted in a distinct livery and hence never service other routes. For Route 31 it did occur frequently and for that reason, in combination with the fact that there were roadworks on the A772 during the observation period, this route will not be further considered in this paper.

In all datasets, some measurements appeared to be erroneous or the result of extreme conditions. For example, gaps have been observed in the dataset, possibly caused by temporary connection losses. This could lead to extremely small measurements if the connection was restored right before a patch was exited. Hence, a data cleaning operation was performed in which measurements were discarded if they were over the median plus three times the standard deviation in that patch, or if they were below 30 seconds (the time needed to clear the smallest patch --- Patch~8 of Figure~\ref{fig: patches airlink} --- at 36mph). The impact was limited: for example, 18 out of 742 measurements across all patches were discarded in the east-west direction between 9AM and 3PM. 


\subsection{Related Work}

This paper broadly follows the same approach as in \cite{reijsbergen2015patch}, in which a patch-based analysis was carried out to evaluate the impact of the introduction of trams to Edinburgh's city centre, and \cite{vissat2015finding}, in which Erlang distributions were used to model the time spent by buses between major stops, and to suggest improvements to bus timetables. The contributions of this paper are the use of a probability distribution recommended in the traffic engineering literature \cite{pline1992traffic}, the use of the statistical model checking engine of UPPAAL \cite{larsen1997uppaal} and the application to a new case study that has received considerable media attention in Edinburgh.

The literature on the impact of speed limits does not appear to be very extensive. One approach would be to use of a traffic simulation tool such as the microscopic tool SUMO \cite{behrisch2011sumo} or the mesoscopic formalism of \cite{burghout2005mesoscopic}. Another approach would be the use of statistical analysis of the impact of speed limits in other cities; see \cite{archer2008impact} for a literature overview.



\section{Probabilistic Modelling}
\label{sec: modelling}

In this section, we present the probabilistic models used in the analysis section of this paper. The most important probabilistic concept that we use is that of the exponentially distributed \emph{phase}. In particular, the time spent in each patch will be modelled as a function of several such phases. We begin with a formal discussion of the above, then discuss parameter fitting and the expression of these models using the language of the model checking tool UPPAAL.

\subsection{The Basic Distributions}

Recall that the time $T$ spent in a phase is \emph{exponentially distributed} with rate $\lambda>0$ if its \emph{probability density function} (pdf) is given by
\[
f_T(t;\lambda) = \lambda e^{-\lambda t}
\] 
for $t \geq 0$ and by $0$ otherwise. The expected amount of time spent in a phase is given by $\frac{1}{\lambda}$; hence, large values of $\lambda$ mean that the time spent in a phase is small on average. The pdf $f_T(t)$ of a random variable $T$ determines the probability of observing values from a small interval around $t$, and will be used later in this paper to measure how well a fitted probability distribution corresponds to the data.

If $T_1,\ldots, T_k$ are exponentially distributed random variables with rate $\lambda$, then $S = T_1 + \ldots + T_k$ is \emph{Erlang-distributed} with rate $\lambda$ and shape $k$. Its pdf equals
\begin{equation}
f_S(s;k, \lambda) = \frac{\lambda^k s^{k-1} e^{-\lambda s}}{(k-1)!} \label{eq: erlang pdf}
\end{equation}
for $s \geq 0$ and $0$ otherwise. 

As mentioned in Section~\ref{sec: background}, we will compare the modelling accuracy of two probability distributions used to model the time spent in a patch; both are generalisations of the Erlang distribution. The first is the \emph{hyper-Erlang distribution}, which is constructed in the following way. Given some $m \in \mathbb{N}$, let $\alpha_1,\ldots,\alpha_m$ be a probability distribution over $\{1,\ldots,m\}$, i.e., $\forall i$, \mbox{$\alpha_i \in [0,1]$} and $\sum_{i=1}^{m} \alpha_i = 1$. Furthermore, let $S_i$, $i \in \{1,\ldots,m\}$, be Erlang-distributed with rate $\lambda_i$ and shape $k_i$ --- these random variables are called the \emph{branches} of the hyper-Erlang distribution. We then say that a random variable $Y$ is {hyper-Erlang-distributed} with shapes $\vec{k} = k_1,\ldots,k_m$, rates $\vec{\lambda} = \lambda_1,\ldots,\lambda_m$ and branch probabilities $\vec{\alpha} = \alpha_1,\ldots,\alpha_m$ if
\begin{equation}
f_{Y} (y; \vec{k}, \vec{\lambda}, \vec{\alpha}) = \sum_{i=1}^m \alpha_i \frac{\lambda^{k_i} x^{k_i-1} e^{-\lambda_i x}}{(k_i-1)!} \label{eq: hyperl pdf}
\end{equation}
for $y \geq 0$, where $f_Y$ denotes the pdf of $Y$.

The second generalisation of the Erlang distribution is one where a real constant $c > 0$ is added. This gives rise to a random variable $Z$ with pdf 
\begin{equation}
f_{Z} (z; k, \lambda, c) = \frac{\lambda^k (z-c)^{k-1} e^{-\lambda (z-c)}}{(k-1)!} \label{eq: three par pdf}
\end{equation}
for $z \geq c$. This is a variation of the probability distribution recommended in the Traffic Engineering Handbook \cite{pline1992traffic}; in their version, the factorial in \eqref{eq: three par pdf} is replaced by a gamma function, thereby allowing $k$ to be any positive real value instead of a positive integer. We require $k$ to be an integer in order to be able to express the distribution in UPPAAL. In the following, we will call this distribution `Erlang+$c$'.

\subsection{Parameter Fitting}

Assuming that we have chosen either the hyper-Erlang or the Erlang+$c$ distribution for the time spent in a patch, the next question is how to choose the distribution parameters such that they correspond to the data in the best possible way. There are several methods for this; examples include the method of moments and maximum likelihood estimation. We use the latter, for which the idea is to maximise the product of the densities in each individual sample --- or rather the logarithm thereof. To make this more concrete, assume we have a sample $\vec{x} = (x_1,\ldots,x_n)$ of observations of the time spent in a patch. The total \emph{log-likelihood} of observing $\vec{x}$ as realisations of $Y$ with parameters $\vec{k}, \vec{\lambda}$, and $\vec{\alpha}$ is then given by
\[
  l_Y(\vec{x};\vec{k}, \vec{\lambda}, \vec{\alpha}) = \sum_{i=1}^n \log(f_Y(x_i;\vec{k}, \vec{\lambda}, \vec{\alpha}))
\]
with $f_Y$ as in \eqref{eq: hyperl pdf}. The log-likelihood $l_Z$ of observing $\vec{x}$ as realisations of $Z$ is defined analogously for parameters $k,\lambda$, and $c$, and $f_Z$ as defined in \eqref{eq: three par pdf}. The functions $l_Y$ and $l_Z$ are then maximised over the available parameters to obtain an optimal fit to the data. For the hyper-Erlang distribution, this is done using the tool HyperStar \cite{reinecke2012hyperstar}, which uses the expectation maximisation algorithm described in \cite{thummler2006novel} and \cite{reinecke2012cluster}. For the Erlang+$c$ distribution, we use a purpose-built numerical optimisation routine based on gradient/steepest ascent (see, for example,~\cite{gonnet2009scientific}) with golden section search. We discuss this in more detail below. 

We first assume that $k$ is given and that we need to find the values of $\lambda$ and $c$ on $\mathbb{R}^2$ that maximise $l_Z$. The (partial) derivatives $\frac{\partial}{\partial\lambda} l_Z$ and $\frac{\partial}{\partial c} l_Z$ are known in closed form, and we use this knowledge to construct an iterative procedure in which we move closer to the nearest optimum in each step. 
First, we initialise $\lambda$ and $c$; with $\bar{x}$ denoting the average of $\vec{x}$ and $\min(\vec{x})$ its minimum, we have found that \mbox{$c = \frac{9}{10} \min(\vec{x})$} and $\lambda = \frac{k}{\bar{x}-c}$ are generally good choices for $\lambda$ and $c$ respectively (the idea behind the factor $\frac{9}{10}$ is that we want $c$ to start close to the smallest observation but not exactly at it). Then, given the values of $\lambda$ and $c$ in the current iteration, we define
\begin{equation}
	h(u) = l_Z(\vec{x};k, \lambda + u \frac{\partial}{\partial\lambda} l_Z(\vec{x}; k, \lambda, c), c + u \frac{\partial}{\partial c} l_Z(\vec{x}; k, \lambda, c)). \label{eq: h function}
\end{equation}
This reduces each iteration to single-dimensional optimisation of $h$, as $h$ is simply $l_Z$ defined on a line in the direction of the steepest ascent. Note that $u=0$ corresponds to $\lambda$ and $c$ remaining the same as in the previous iteration, so we search for a non-zero value of $u$ that maximises (or at least improves) $h(u)$. When this value has been found, the next iteration's values of $\lambda$ and $c$ can be computed as $\lambda + u \frac{\partial}{\partial\lambda} l_Z(\vec{x}; k, \lambda, c)$ and $c + u \frac{\partial}{\partial c} l_Z(\vec{x}; k, \lambda, c)$ respectively. 

The function $h$ equals $-\infty$ in parts of $\mathbb{R}$; this happens when the corresponding value of $\lambda$ is negative or when $c$ becomes either negative or larger than the smallest observed value. This proves problematic for all of the general-purpose optimisation procedures implemented in R through the function \mbox{{\tt optim}}, including Brent's method which is specifically intended for single-dimensional optimisation. Accordingly, we have implemented our own algorithm which follows the principles of golden section search. We search for the local optimum of $h$ on $[-1, 1]$, which means that in each iteration we move away from the old values of $\lambda$ and $c$ by at most the gradient (to see this, substitute $-1$ or $1$ into \eqref{eq: h function}). Then, we fix an integer $l \in \{2,3,\ldots\}$ to be used throughout the entire procedure, and compute the values of $h$ in the points $-1, -1+\frac{1}{l}, -1+\frac{2}{l}, \ldots, 1$. We then determine the smallest value $i \in \{2,3,\ldots, 2l\}$ for which $h(\frac{i-1}{l}-1)  > h(\frac{i}{l}-1)$. The reason is that we know that the value of $u$ that maximises $h(u)$ must then be on $[\frac{i-2}{l}-1, \frac{i}{l}-1]$ --- after all, the derivative of $h$ can only change sign once if we assume that a global maximum exists. We found that $l=2$ does not necessarily give the best run times; we used $l=7$ throughout. We run the single-dimensional optimisation procedure several times (in our implementation, we took ten steps), then choose $u$ to be the centre of the remaining interval and complete the iteration. We continue until the changes in terms of $\lambda$ and $c$ have dropped below a fixed threshold or if a fixed number of iterations has been reached.

Since we now have a procedure for finding the optimal values of $\lambda$ and $c$ as a function of $k$ and $\vec{x}$, the final challenge is to find the optimal $k$. We use the fact that $k$ is discrete by starting with $k=1$ and then increasing $k$ until $l_Z$ starts to decrease. In our experiments, this appeared to return the global optimum. As can be seen in the `Erlang+$c$' column of Table~\ref{tab: phase counts}, the optimal values of $k$ were typically not very large, so it was not necessary from a computational point of view to find a more efficient way of initialising this procedure.


\subsection{Probabilistic Timed Automata and UPPAAL}


UPPAAL \cite{larsen1997uppaal} is a toolbox for the analysis of real-time systems modelled as \textit{timed automata}. Of particular interest to this paper is the UPPAAL Statistical Model Checker (SMC) \cite{david2011time}, which allows for the verification of Probabilistic Timed Automata (PTAs), the stochastic extension of timed automata. We do not give a full description of PTAs but discuss the key language features used in the paper through a discussion of the PTA models for the patch-based model. 


\begin{figure}[!h]
	\centerline{\includegraphics[width=0.5\textwidth]{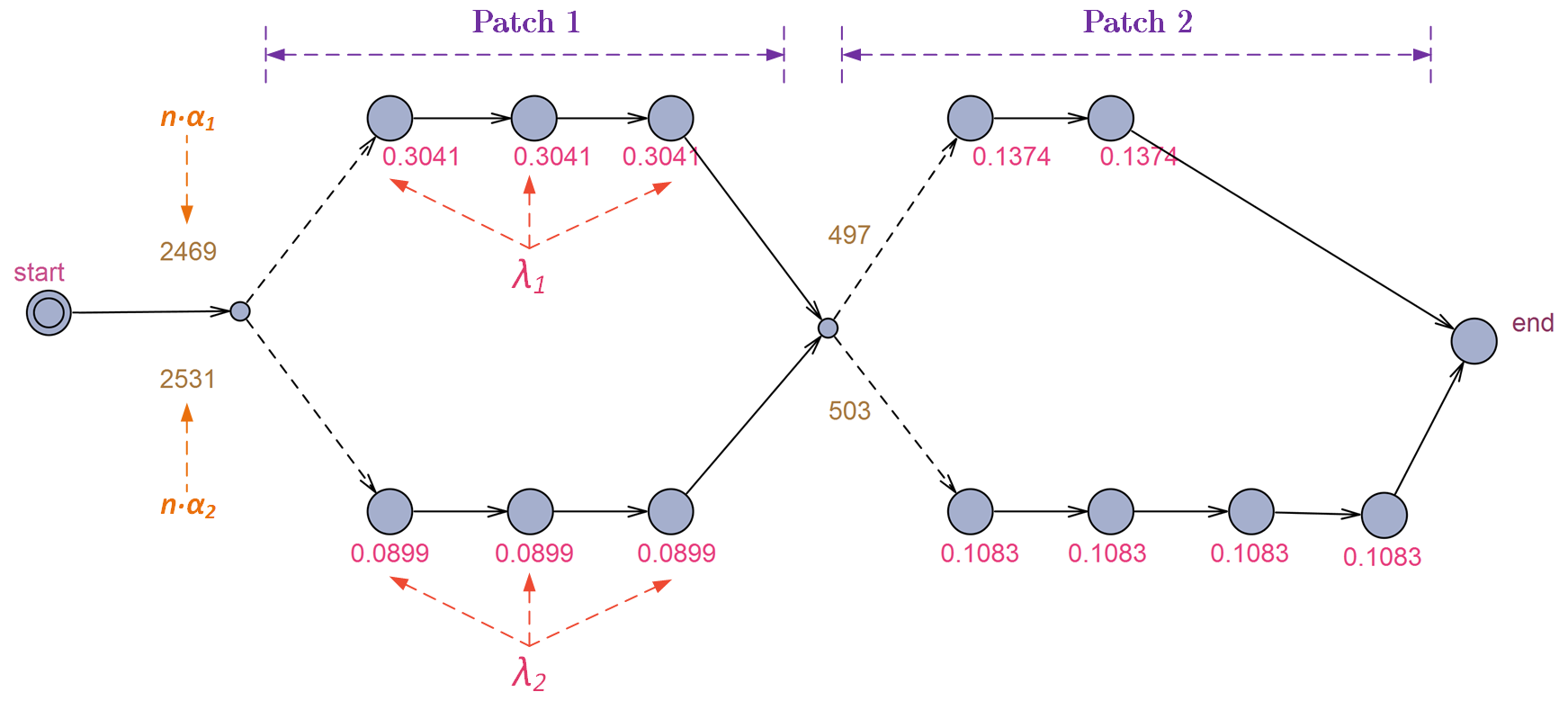}}
    \caption{Illustration of a PTA representing a system with two patches with two-branch Hyper-Erlang distributed crossing times.}
    \label{fig:Uppaaldemo1}
\end{figure}

In Figure~\ref{fig:Uppaaldemo1}, we depict a sequence of two patches for which the journey time is modelled using a two-branch hyper-Erlang distribution. States (called `locations' in PTA terminology) are represented using circles, and edges (representing transitions between states) using arrows. The system starts in the left-most location (the double circle), which is exited with a rate of $1$ second (this transition would ideally be immediate, but the 1-second delay is negligible). We need this state because the system requires the specification of a unique initial state. When the initial state is left, a probabilistic choice is made between the two branches of the next patch. The edge weights, which determine the relative likelihood of choosing an edge, are required to be integers. Hence, the values of $\alpha_1$ and $\alpha_2$ --- found to be $0.4938$ and $0.5062$ by the parameter-fitting algorithm --- had to be converted to the smallest pair of integers with the same ratio, which in this case were 2469 and 2531. 

If branch $i$ has been chosen, the system then needs to complete $k_i$ locations that each have rate $\lambda_i$. When this has been completed, the system jumps to a so-called branch point (the small circle) from which a probabilistic choice is made between the branches of the next patch. When the final patch has been completed, the system jumps to the final state labelled `end'.

\begin{figure}[!h]
	\centerline{\includegraphics[width=0.5\textwidth]{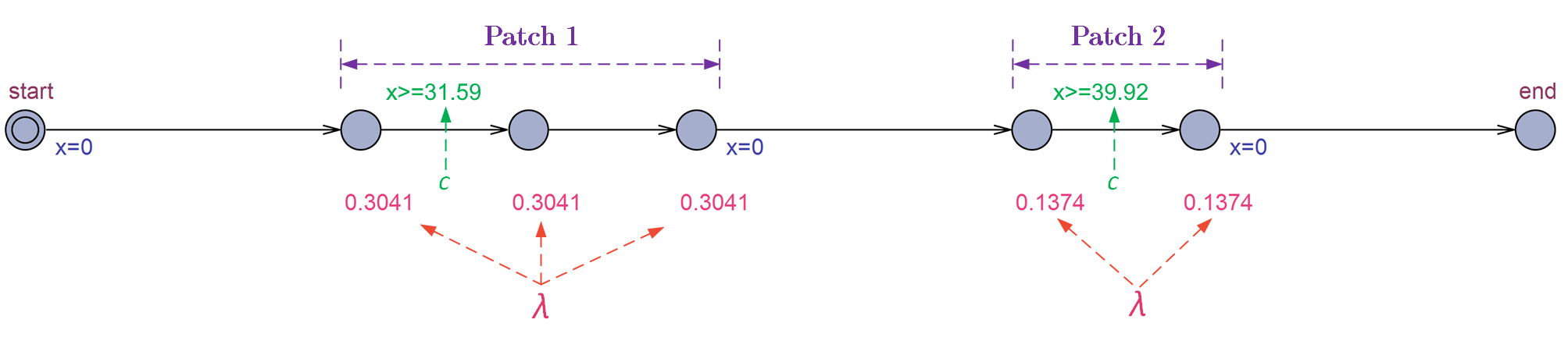}}
    \caption{Illustration of a PTA representing a system with two patches with Erlang+$c$ distributed crossing times.}
    \label{fig:Uppaaldemo2}
\end{figure}

In Figure~\ref{fig:Uppaaldemo2}, we depict a similar model for the Erlang+$c$ distribution. In addition to the discrete states, we now also use a real-valued clock, called $x$. This clock is reset immediately after a patch is left. We then impose a guard on the first (or only) edge of the patch requiring that this edge cannot be completed before the clock $x$ has reached a value greater than $c$; in pratice, this adds a constant delay of $c$ to the global clock. Otherwise, the behaviour in each patch is similar to that of a single Erlang branch in Figure~\ref{fig:Uppaaldemo1}.

Whereas Figures~\ref{fig:Uppaaldemo1}~and~\ref{fig:Uppaaldemo2} depict graphical representations of the PTA models used in this paper, the same models can also be expressed in XML format. These XML files can be generated by R, taking as input either a file with the hyper-Erlang parameters obtained via HyperStar, or a matrix of the Erlang+$c$ distribution parameters obtained by another function in R. Furthermore, the system property of interest is also embedded in the XML file. We are interested in two probabilities: the property of a bus reaching the final location over one minute early or less than five minutes late.\footnote{The notion of a bus being `on-time' if it is at most $N_1$ minutes early and at most $N_2$ minutes late is a common notion, and the particular choice of $N_1=1$ and $N_2=5$ is from the Scottish Government's Bus Punctuality Improvement Partnerships guiding document \cite{bpips}. This notion of punctuality is less relevant for the Airlink as it is a `frequent' service during the day, meaning that different standards apply; however, it is infrequent at night. } For example, if the timetabled journey duration to get from Waverley station in Edinburgh's city centre to the airport is 28 minutes (i.e., 1680 seconds) then the property
\[
	\textbf{\textit{\text{Pr}[<=1620] (<> \text{Process.end})}}
\]
specifies the property of being over a minute early (`Process' is the name of the system and `end' the name of the final state). The SMC engine of UPPAAL is then invoked to estimate the probability of this property being satisfied. In particular, system executions are repeatedly executed and the results used to make statistically justified statements about the relevant probabilities. 



%

\section{Analysis \& Results}
\label{sec: results}

In this section, we present the results of applying the methodology described above to the case study of the proposed speed limit in Edinburgh. The section is divided into two parts. In Section~\ref{sec: modelling results}, we discuss the results of the parameterisation and model construction steps, and discuss the impact of the modelling choices. In Section~\ref{sec: speed limit} we discuss the actual impact of the speed limit using the model built in the previous step.

\subsection{Modelling Results} \label{sec: modelling results}

In order to best illustrate the impact of the modelling choices, we focus on a specific dataset of GPS measurements, namely the east-west direction of the Airlink and the time window between 9AM and 3PM, without the speed limit. This dataset consists of 724 measurements with 68-76 measurements per patch. The differences between the numbers of observations was due to the outlier removal; outliers occurred more often in the patches near the two ends of the route and less so in the centre.

\begin{table}
	\centering
		\begin{tabular}{ccccc}
		  \hline
			 & & & \multicolumn{2}{c}{hyper-Erlang} \\
			Patch & Erlang+$c$ & Erlang & 2-branch & 3-branch\\
			\hline
			
			
			1 & -338 & -343 & -340 & -337 \\
			2 & -346 & -386 & -296 & -286 \\
			3 & -336 & -336 & -336 & -334 \\
			4 & -357 & -358 & -357 & -356 \\
			5 & -396 & -401 & -396 & -394 \\
			6 & -333 & -339 & -336 & -334 \\
			7 & -358 & -361 & -356 & -356 \\
			8 & -409 & -409 & -409 & -409 \\
			9 & -388 & -388 & -387 & -384 \\
			10 & -447 & -449 & -442 & -440 \\

			\hline
		\end{tabular}
		\caption{Total log-likelihoods for the distributions fitted to the patch journey times in each of the patches for the Airlink, from Waverley Bridge to the airport during midday, rounded to integers. }
		\label{tab: log-liks}
\end{table}

In Table~\ref{tab: log-liks}, we display for each patch and for four choices of distributions (Erlang, 2-branch hyper-Erlang, 3-branch hyper-Erlang, and Erlang+$c$) the log-likelihood of observing the data under the distribution with the parameters that were found to be optimal. As can be seen form the table, the numbers tend not to differ starkly. 


In nearly all cases the 3-branch hyper-Erlang has a higher log-likelihood than the Erlang+$c$. However, this comes at a price: the number of phases tends to be much smaller for the Erlang+$c$ than for the other distributions. In Table~\ref{tab: phase counts}, we display the expected number of completed phases: this is just $k$ for the Erlang and Erlang+$c$, and $\sum_i \alpha_i k_i$ for the hyper-Erlangs. The differences are considerable, sometimes even between the Erlang and Erlang+$c$ (see, for example, Patch~1). The consequences are twofold. Firstly, since the time to draw a sample using the SMC engine of UPPAAL scales roughly linearly with the expected number of completed phases, it takes less time to obtain accurate estimates of the on-time probabilities using the Erlang+$c$. Secondly, high phase counts can be a symptom of overfitting: it could indicate that a sharp peak around a cluster of observations has been created through an Erlang branch with many phases and a rate such that the branch mean is near the centre of the cluster.

\begin{table*}[!htbp]
\centering
\small
\noindent\makebox[\textwidth]{
\begin{tabular}{c c c c c c c c}
  \hline 
	& & & \multicolumn{2}{c}{Speed Limit} & \\
Route & {Period} & {Direction} & Timetable & Old & New & Difference & Perc. Change \\
  \hline
\multirow{4}{*}{Airlink} & \multirow{2}{*}{9AM - 3PM } & Waverley-Airport & 1680 & 1886.34 & 1896.82 & 10.48 & 0.56\% \\
 &  & Airport-Waverley & 1800 & 2038.05 & 2045.66 & 7.61 & 0.37\%\\
 & \multirow{2}{*}{12AM - 5AM } & Waverley-Airport & 1260 & 1413.83 & 1429.01 & 15.18 & 1.07\% \\
 & & Airport-Waverley & 1260 & 1208.13 & 1232.43 & 24.30 & 2.01\% \\[0.1cm]
N11 & \multirow{1}{*}{12AM - 5AM } &  Ocean Terminal-Hyvots Bank & 2460 & 2256.93 & 2289.27 & 32.34 & 1.43\% \\
   \hline
\end{tabular}
}
\caption{We depict the average total journey times in seconds both with and without incorporation of the the speed limit impact as per \eqref{eq: speed limit}. This is done for two routes, the Airlink in both directions and during two different time periods, and for the N11 from Ocean Terminal to Hyvots Bank (the N11 only goes in one direction). Timetable predictions were obtained using the journey planner on \url{http://lothianbuses.com/getting-around/journey-planner}.} 
\label{table:means}
\end{table*}  

\begin{table}[!htpb]
	\centering
		\begin{tabular}{ccccc}
		  \hline
			 & & & \multicolumn{2}{c}{hyper-Erlang} \\
			Patch & Erlang+$c$ & Erlang & 2-branch & 3-branch \\
			\hline
			1 & 4 & 80 & 149.5 & 591.4 \\
			2 & 1 & 3 & 186.7 & 310.8 \\
			3 & 13 & 32 & 59.0 & 569.9 \\
			4 & 3 & 10 & 36.4 & 39.9 \\
			5 & 3 & 13 & 47.5 & 89.0 \\
			6 & 2 & 38 & 125.5 & 161.6 \\
			7 & 3 & 23 & 79.6 & 91.4 \\
			8 & 5 & 7 & 13.2 & 21.8 \\
			9 & 9 & 9 & 22.0 & 272.9 \\
			10 & 6 & 12 & 13.5 & 27.8 \\
			\hline
		\end{tabular}
		\caption{The expected number of exponential phases for each of the distributions.}
		\label{tab: phase counts}
\end{table}

\begin{figure*}[!htbp]
	\centering
	\subfloat[][Erlang+$c$]{\includegraphics[width=0.44\textwidth]{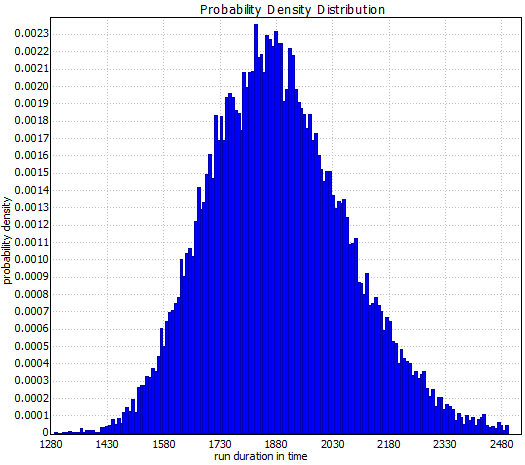}\label{fig:roads}}
	\hspace{1cm}
  \subfloat[][2-branch hyper-Erlang]{\includegraphics[width=0.44\textwidth]{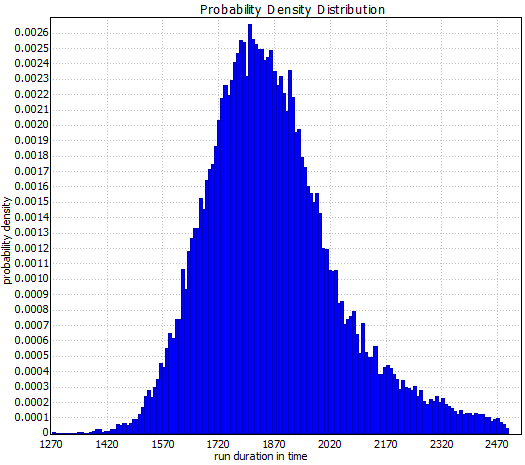}\label{fig:pie}}
	\caption{Observation histograms for the time needed to complete the entire Waverley-airport journey based on a model in which each patch either has an Erlang+$c$ or a 2-branch hyper-Erlang distribution, generated using UPPAAL. The hyper-Erlang distributions result in a stronger skew towards higher values.}
	\label{fig_ uppaal hists}
\end{figure*}

The choice of distribution does have a big impact on the tail of the distribution of the time needed to complete the entire journey. This is displayed in Figure~\ref{fig_ uppaal hists}, in which it can be seen that the hyper-Erlang distributions have a noticeably heavier tail on the right. As mentioned earlier, the time needed to generate samples for the Erlang+$c$ model is lower than for the hyper-Erlang model; whereas Figure~\ref{fig_ uppaal hists}(a) is based on 25,103 samples generated in 6 seconds, Figure~\ref{fig_ uppaal hists}(b) is based on 22,116 samples generated in 39 seconds. We found that the distribution of total journey times does seem to have a relatively heavy tail, which would support the use of the hyper-Erlang distribution.

Finally, we consider the impact of the modelling choice on the computation of the on-time probabilities, which we will discuss in more detail below. As can be seen in Table~\ref{table:probs}, the choice of distribution can have a big impact, in our case even more so than the speed limit itself. 


\subsection{Patch Parameter Fitting Results}

To better illustrate the results of fitting the hyper-Erlang distribution to the individual patch sojourn time datasets, we provide screen shots taken of the HyperStar tool in Figures~\ref{fig: time dists a}~and~7. These screen shots depict the datasets of the time Airlink buses spent in each of the ten patches of the Airlink route (see Figure 2), both in terms of an observation histogram and the empirical Cumulative Distribution Function (CDF). We only consider buses heading from Waverley Bridge to the airport between 9AM and 3PM. In the same figure as the histograms and empirical CDFs, we depict the Probability Density Functions (PDFs) and CDFs of the hyper-Erlang distributions fitted to the data. For each dataset, we consider three types of hyper-Erlang distributions: single-branch (i.e., Erlang distributions), 2-branch, and 3-branch. The coloured vertical lines in the graphs depict the manually selected peaks which determine the maximum number of branches and inform the choice of initial parameters for the implemented parameter optimisation procedure (the EM algorithm).

As can be seen in the figures, the dataset for which the hyper-Erlang does best when compared the the Erlang$+c$ distribution is Patch 2, which appears to contain quite a few outliers. We mention again here that we do apply some form of outlier removal, but the used criterion for removal --- namely a large distance from the mean relative to the standard deviation --- does not lead to good results when the number of outliers is so high that the standard deviation becomes too large. Something similar applies to Patch 10, which has outliers in the tail on the left. The improvement of Erlang$+c$ over standard Erlang is the biggest for Patches 5 and 6. In general, the hyper-Erlang distributions have a tendency to overfit peaks in the data, which is particularly noticeable for Patch 9 and 3 Erlang branches.

\subsection{The Impact of the Speed Limit} \label{sec: speed limit}

\begin{table*}[!htbp]
\centering
\small
\begin{tabular}{c c c c c}
  \hline 
	Route & Distribution & Speed Limit & Too Early & Too Late \\
  \hline
	\multirow{4}{*}{ Airlink } & \multirow{2}{*}{ Hyper-Erlang } & old & [0.091799, 0.092799] & [0.193120, 0.194120] \\  
 & & new & [0.088044, 0.089044] & [0.210440, 0.211440] \\ 
 & \multirow{2}{*}{ Erlang+$c$ } & old & [0.073240, 0.074240] & [0.335775, 0.336775] \\ 
 & & new & [0.067363, 0.068363] & [0.345214, 0.346214]\\
   \hline
	

\end{tabular}
\caption{90\% confidence intervals for the probability that a bus completes the Waverley-Airport journey during the 9AM-3PM period over one minute too early (so before 1620 seconds) or over five minutes too late (so after 1980 minutes). The hyper-Erlang distributions had two branches. All distributions were fully parameterised again for the all patches, even those which would be unaffected by the speed limit.
} 
\label{table:probs}
\end{table*}

To evaluate the impact of the speed limit, we create a second dataset in the following way: we test for each GPS measurement (corresponding to a 5-second period) whether the recorded speed of the vehicle $s$ was over 20mph, and if so, we add the value 
\begin{equation}
\frac{s - 20}{20} \cdot 5 \label{eq: speed limit}
\end{equation} 
to the total time spent in the corresponding patch. This quantity represents the additional amount of time needed to complete the same distance under the speed limit. It allows us to parameterise the model using both datasets and compare expected journey times and on-time probabilities. Note that \eqref{eq: speed limit} of course represents a rough approximation: if, for example, the speed limit also changes the density of vehicles on the road, then this could have an additional impact on bus travel times. One way of investigating this effect would be to use a microscopic road simulation tool such as SUMO, which incorporates both the buses and the other road traffic as individual vehicles in the simulation. However, since we do not have access to reliable non-bus traffic data, we are currently unable to realistically parameterise such a model, so this is left as a direction for future research.

As can be seen in Table~\ref{table:means}, the impact of the speed limit varies across time periods and routes. For the Airlink, which is only impacted by the speed limit on the segment between Waverley Station and Haymarket (Patches 8, 9 and 10 of Figure~\ref{fig: patches airlink}), the impact on the timetable seems to be negligible, as buses only rarely manage to go over 20mph in the city centre. As expected, the impact at night is larger than during the day. The N11, which is impacted by the speed limit on over half its route, is slowed down by about 58.91 seconds, although the total impact is still limited relative to the total route duration (namely about 2.27\%). We do note that we observed that some N11 buses switched routes during the night, and although this did not appear to have a substantial impact the results are subject to this caveat.


The impact of the speed limit also does not seem to have much of an effect on the probabilities of not being on-time for the Airlink, as can be seen in Table~\ref{table:probs}. However, for the N11, the full minute being added on to the average route completion did have a noteworthy impact. It should be noted that buses of all routes typically spend a large amount of time stationary, particularly to allow passenger to board or alight, which mitigates the effect of the speed limit.

\section{Conclusions}
\label{sec: conclusions}

In this paper we have presented a probabilistic model that was able to quantify the impacts on bus timetables of the newly proposed 20mph speed limit plan in Edinburgh. We considered two ways of incorporating the stochastic nature of bus movement within route segments (`patches'), namely the use of hyper-Erlang distributions and an Erlang distribution plus a constant that is a variation of the distribution recommended by the Traffic Engineering Handbook. The tail behaviour of the former is better able to represent total journey durations, whereas the latter is more efficient to simulate and its parametrisation is more stable across repeated experiments. We combined the patches into a single representative model in UPPAAL which allowed us to compute interesting punctuality metrics. The impact of the speed limit was found to be limited.

Regarding future work, patch selection and dimensioning is a time-consuming non-trivial task that easily becomes cumbersome when the number of patches and routes is increased. Automated patch selection would greatly enhance the efficiency of our methodology. We did try to find patch centres using the $k$-means clustering algorithm, however one of the disadvantages is that there is no easy way to align patch boundaries to the roads affected by the speed limit. As such, we would need to find a way to automatically incorporate the map data of the proposed speed limit.

Also, the speed limit adjustment as given in Equation \eqref{eq: speed limit} is of course an approximation, and assumes that traffic conditions remain the same. A micro-scale simulation model could be constructed to validate this assumption. However, since we do not have access to full traffic data it would be a challenge to realistically parameterise such a model.

\section{Acknowledgments}

This work has been supported by the EU project QUANTICOL, 600708.
The authors thank Bill Johnston and Philip Lock of Lothian Buses for providing access to the data and for their helpful feedback on parts of this research project, and Jane Hillston for her helpful comments.


\sloppy

\bibliographystyle{plain}
\bibliography{2015_valuetools}


\newcommand{\patcha}[3]{\subfloat[#1][Patch #2, #3 \ifboolexpr{ test {\ifnumgreater{#3}{1}} } {branches} {branch}.]{\includegraphics[width=0.28\textwidth,trim={1450px 309px 372px 130px},clip]{#2_#3.png}}}
\newcommand{\patchb}[3]{\subfloat[#1][Patch #2, #3 \ifboolexpr{ test {\ifnumgreater{#3}{1}} } {branches} {branch}.]{\includegraphics[width=0.28\textwidth,trim={1452px 309px 340px 130px},clip]{#2_#3.png}}}

\newgeometry{a4paper,bindingoffset=0.2in,%
            left=0.5in,right=0.5in,top=0.5in,bottom=0.5in,%
            footskip=.25in}

\begin{figure*}[p]
\captionsetup[subfigure]{labelformat=empty}
\centering
\patcha{0}{1}{1} \patchb{1}{1}{2} \patchb{2}{1}{3} \\
\patchb{3}{2}{1} \patchb{4}{2}{2} \patchb{5}{2}{3} \\
\patcha{6}{3}{1} \patcha{7}{3}{2} \patcha{8}{3}{3} \\
\patchb{9}{4}{1} \patchb{10}{4}{2} \patcha{11}{4}{3} \\
\patcha{12}{5}{1} \patcha{1}{5}{2} \patchb{2}{5}{3}
\caption{Hyper-Erlang parameter fitting results for Patches 1-5 of the Airlink route.}
\label{fig: time dists a}
\end{figure*}

\begin{figure*}[p]
\captionsetup[subfigure]{labelformat=empty}
\centering

\patcha{15}{6}{1} \patchb{4}{6}{2} \patcha{5}{6}{3} \\
\patchb{18}{7}{1} \patcha{7}{7}{2} \patcha{8}{7}{3} \\
\patcha{21}{8}{1} \patchb{1}{8}{2} \patchb{2}{8}{3} \\
\patchb{24}{9}{1} \patcha{4}{9}{2} \patcha{5}{9}{3} \\
\patcha{27}{10}{1} \patcha{7}{10}{2} \patchb{8}{10}{3}
\caption{Hyper-Erlang parameter fitting results for Patches 6-10 of the Airlink route.}
\label{fig: time dists b}
\end{figure*}

\restoregeometry

\end{document}